\documentclass[%
 reprint,
 amsmath,amssymb,
 aps,
]{revtex4-1}

\usepackage{graphicx}
\usepackage{dcolumn}
\usepackage{bm}

\usepackage{amsmath}

\usepackage{color}

\begin{document}

\preprint{APS/123-QED}

\title{Dark Energy Stars with Phantom Field}

\author{Muhammad F. A. R. Sakti$^{1,2}$}
\email{fitrahalfian@gmail.com}
\author{Anto Sulaksono$^{3}$}%
 \email{anto.sulaksono@sci.ui.ac.id}
\affiliation{$ ^1 $Theoretical Physics Laboratory, THEPi Division,}
\affiliation{$ ^2 $Indonesia Center for Theoretical and Mathematical Physics (ICTMP), Institut Teknologi Bandung, Jl. Ganesha 10 Bandung, 40132, Indonesia,}
\affiliation{$ ^3 $Department of Physics, Universitas Indonesia, Depok, Indonesia 16424.}

%
%
%

\date{\today}

\begin{abstract}
Dark energy is the constituent with an enormous abundance of the present universe, responsible for the universe's accelerated expansion. Therefore, it is plausible that dark energy may interact within any compact astrophysical objects.  The author in Ref. [Phys. Rev. D 83, 127501 (2011)], constructs an exact star solution consisting of an ordinary matter and phantom field from a constant density star (CDS) known as Schwarzschild interior solution. The star denotes a dark energy star (DES). The author claims that the phantom field represents dark energy within the star.  So far,  the role of the phantom field as dark energy in DES is not systematically studied yet. Related to this issue, we analyze the energy condition of DES. We expect that DES shall violate the strong energy condition (SEC) for a particular condition. We discover that SEC is fully violated only when the compactness reaches the Buchdahl limit. Furthermore, we also investigate the causal conditions and stabilities due to the convective motion and gravitational cracking. We also find that those conditions are violated. These results indicate that DES is not physically stable. However, we may consider DES as an ultra-compact object of which we can calculate the gravitational wave echo time and echo frequency and compare them to those of CDS. We find that the contribution of the phantom field delays the gravitational wave echoes. The effective potential of the perturbed DES is also studied. The potential also enjoys a potential well like CDS but with a deeper well. We also investigate the possibility that DES could form a gravastar when $ C=1 $. It is found that gravastar produced from DES possesses no singularity with \textcolor{black}{a dS-like} phase as the interior. These results could open more opportunities for the observational study of dark energy in the near future, mostly from the compact astrophysical objects.
\begin{description}
\item[PACS numbers]
\end{description}
\end{abstract}

\maketitle


\section{Introduction}

It is well known that the expansion of the present universe is accelerated, not slowing down due to gravity as expected before. The observational results prove an unknown nature that governs this accelerated expansion \cite{PerlmutterAstroJ}. This nature is called as dark energy. From the observations, the dark energy fills up around 70\% of the universe. Dark energy possesses an unusual property, i.e.,  negative pressure. This negativity corresponds to the violation of the strong energy condition (SEC). Several remarkable dark energy models are known in literature such as cosmological constant, quintessence, K-essence, phantom field, tachyon field, dilatonic field, and Chaplygin gas \cite{Copeland2006}.
   
Since dark energy fills up most of our universe, this matter may also interact with any ordinary matter within any local astrophysical manifestations such as wormholes, black holes, compact stars, and other exotic compact objects. The equation of state (EoS) of dark energy can be described by $ p=\omega  \rho $ with $ \omega <-1/3 $. The first black hole solution with such an EoS is constructed in \cite{KiselevCQG2002} and called as a quintessential black hole. The extended solutions of this quintessential black hole and their studies including the quantum aspects can be found in Refs. \cite{FernandoGRG2012,ToshmatovAhmedovEPJP2017,Wang2017,GhoshEPJC2016,SaktiarXivString2019,Chen2008,SaktiEPJP,HadyanIJMPD2020,SaktiAnnPhys2020,SaktiZenarXiv2020}. One can find the study of such EoS within the stars at Refs. \cite{BharRahamanEPJC2015,LoboCQG2006}. Besides, wormholes supported by dark energy are also studied in \cite{BharRahamanEPJC2016} of which the authors take $ \omega <-1 $ to portray phantom dark energy. Although the EoS can be formulated in the form $ p=\omega  \rho $, as long as SEC is still violated, such compact objects may have different EoSs but with negative principal pressures. For instance, in Ref. \cite{HorvatCQG2013}, they investigate the boson star with non-minimal coupling to the gravity that yields negative pressures. In Ref. \cite{BertolamiPRD2005}, the dark star is studied with the Chaplygin gas EoS.

One of the intriguing directions in the study of dark energy is to delve into the correct model of dark energy. Phantom field appears in several theories described as the fields whose the kinetic term has flipped sign, yielding negative kinetic energy. In string theory, 
the phantom field emerges in the study of anti-branes \cite{Vafabrane} or ghost branes \cite{OkudaTakayanagiJHEP2006}. It also appears in the Einstein-Maxwell-dilaton system where the dilatonic kinetic term is flipped to be negative \cite{ClementFabrisPRD2009}. Phantom field actually may lead to the emergence of quantum instabilities \cite{CaldwellPLB2002,CaiSaridakisPR2010}. However, the quantum instabilities can be omitted by considering the phantom scalar field, which arises from an effective field theory resulting from a fundamental theory with a positive energy \cite{PiazzaTsujikawa2004}. Consequently, the phantom scalar field has an entirely physical grounding. It can be considered as a candidate for the dark energy model.

In this paper, we study an interior solution constructed in \cite{YazadjievPRD2011} where the phantom field is governed as a dark energy model that interacts minimally with the ordinary matter. Therein, the interior solution of the Einstein field equation with a phantom scalar field is derived by applying a generating method presented in Ref. \cite{YazadjievModPhysLettA} from the Schwarzschild interior solution or the constant energy density star (CDS), namely dark energy star (DES). There are several similarities and significant differences between DES and CDS, which we can be seen in Ref. \cite{YazadjievPRD2011}. Both stars respect the Buchdahl inequality $ C\leq 4/9 $ and have asymptotically flatness at infinity. However, DES possesses a coordinate-dependent energy density and a non-vanishing charge known as a dark charge from the phantom field's contribution. This dark charge has a critical role in determining compactness. Due to the phantom field's presence, the star's pressures are anisotropic, where the phantom field's kinetic term appears on the radial pressure. Another exciting feature of DES is the vanishing tangential pressure when $ |D| \rightarrow M $ while keeping $ R $ fixed.

We know that CDS is the simplest exact interior solution of the Einstein field equation supported by an isotropic perfect fluid. CDS is also used to mimic the Schwarzschild black hole in the limit $ R \rightarrow 2m $ \cite{KonoplyaPRD2019} and resembles the main features of the simplest gravatar model proposed in Refs. \cite{MazurMottolaPNAS2004,MazurMottolaCQG2015}. From this fact, we may expect that DES presented in \cite{YazadjievPRD2011} will be able to mimic a black hole consisting of a phantom dark energy that we call a phantom black hole. Furthermore, DES can also \textcolor{black}{be used to} study the gravastar-like solution with dark energy represented by the phantom field. Due to the constancy of the energy density, CDS violates the causal conditions, even though it is stable due to the convective motion, which satisfies $ \rho ''=0 $ and due to the gravitational cracking since the pressure is isotropic. However, the instability of the isotropic structure can also occur in compact objects.  The stability of the anisotropic structure of neutron stars within the Eddington-inspired Born-Infeld (EiBI) theory has been investigated in Ref. \cite{DanariantoSulaksonoPRD2019}. The primary analysis of those stabilities is given in \cite{HectorEPJC2018}. However, the stabilities due to those conditions for DES are not systematically investigated, yet \cite{YazadjievPRD2011}. Moreover, the claim that the phantom field plays a role as dark energy is not thoroughly investigated \cite{YazadjievPRD2011} where the dark energy shall violate SEC \cite{ChanSilvaMPLA2009}. Indeed, stars which do not contain dark energy shall satisfy all energy conditions, for example, the anisotropic neutron stars \cite{SetiawanSulaksonoEPJC2019} and strange stars in non-conserved energy-momentum theory such as Rastall gravity \cite{MaulanaSulaksonoPRD2019}.

CDS can be used as an illustrative model for exotic compact stars which can produce gravitational echoes \cite{PaniFerrariCQG2018}, even though this represents an unphysical star because of the violation of the causal conditions.  If CDS has compactness in the range of $ 1/3 \leq C \leq 4/9  $, it can have a photon sphere or the light ring where the radiation can be effectively trapped between the light ring the center of the star. Since DES possesses a similar Buchdahl inequality, it may also have the light ring in the ultra-compact region like CDS. Because the dark charge also determines the compactness of DES, it will be interesting to analyze its impact on the gravitational echoes. Moreover, the authors in Ref. \cite{HoughtonarXiv} introduce that the information of dark energy can be accommodated from the gravitational radiation of binary systems of supermassive black holes. However, the feasibility is still needed further investigation (see Ref. \cite{EnanderMortsellPLB2010} and the references therein for \textcolor{black}{detailed explanation}). The gravitational echoes can be detected in the near future with the gravitational waves detectors, so investigating the impact of the phantom dark energy on the gravitational echoes may be possible to probe dark energy locally in DES or phantom black holes.

In Ref. \cite{ConklinHoldomPRD2018}, the authors investigate the gravitational echoes for different exotic compact objects including gravastars given in \cite{MazurMottolaPNAS2004,MazurMottolaCQG2015}. Gravastar in Refs. \cite{MazurMottolaPNAS2004,MazurMottolaCQG2015} is the simplest model of gravastars constructed from CDS which possesses three different regions, i.e interior ($ 0\leq r <R_0 $), thin shell ($ R_0\leq r <R $) and exterior ($ R <r $). Gravastar suffers a discontinuity in EoS on the surface; however, it can be resolved by the discontinuity of the extrinsic curvature by employing Darmois-Israel's junction conditions \cite{Darmois,IsraelNuoCim1966}. Since CDS is the seed solution of DES, we may expect that the gravastar-like solution can also be constructed from DES.

In the present paper, we first investigate the existence of the dark energy properties coming from the phantom field in DES by employing the energy conditions analysis. As we know, CDS does not meet the criteria of causal conditions, so we also check whether DES satisfies these conditions by analyzing the speed of sound in the star. Furthermore, it is also interesting to examine the stabilities of DES due to the convective motion and gravitational cracking because the anisotropy from the phantom field's kinetic term may cause instability within the star. The surface redshift of DES will also be studied to see the phantom field's influence compared to that of CDS. Since we know that CDS is a good toy model for investigating the gravitational echoes for an ultra-compact object, we also delve into the properties of DES as an ultra-compact \textcolor{black}{object}. We compare our results with CDS to see the phantom field's impact on the gravitational echoes, especially in the echo time and echo frequency. We also compare the effective potential between CDS and DES to see how this corresponds to the echo time and echo frequency. Then we examine the gravastar-like solution of DES. We wish to see whether DES can be constructed to be the ordinary gravastar given in ref. \cite{MazurMottolaPNAS2004} or exceptional gravastar. Note that we use the convention $ G=c_l =1 $ for the whole paper. We refer $ c_l $ as the speed of light.

This paper is organized as follows. In Sec. \ref{sec2}, we review
DES as derived in \cite{YazadjievPRD2011}. In Sec. \ref{sec3}, we investigate the energy conditions, stabilities, and surface redshift analysis. In Sec. \ref{sec4}, the properties of an ultra-compact DES are studied, including the calculation of echo time and echo frequency and the analysis of the effective potential. Before the conclusions, we provide the gravastar-like solution of DES in Sec. \ref{sec6}. Finally, in conclusion, we summarize the results of the whole paper.

\section{Dark Energy Star Solution}
\label{sec2}

It is calculated by Yazadjiev \cite{YazadjievPRD2011} using his method in \cite{YazadjievModPhysLettA} that one can obtain an interior solution of the Einstein field equation containing an ordinary matter and a scalar phantom field with no phantom potential. The equation of motion of this system is described by
\begin{equation}
R_{\mu\nu} = \kappa \left(T_{\mu\nu} -\frac{1}{2}g_{\mu\nu}T \right) - 2 \partial_\mu \varphi \partial_\nu \varphi,  \label{eq:EinsteinPhantomEq}
\end{equation}
where $ \kappa=8\pi $ and $ T_{\mu\nu} $ is the energy-momentum tensor of the ordinary matter described by
\begin{equation}
T_{\mu\nu} = (\rho + p)u_\mu u_\nu + pg_{\mu\nu}.
\end{equation}
The phantom kinetic term is given on the second term of right hand side of Eq. (\ref{eq:EinsteinPhantomEq}). $ \rho$ and $ p $ are energy density and pressure of the isotropic perfect fluid, respectively. The equation of motion of the phantom field is given by
\begin{equation}
\nabla_\mu \nabla^\mu \varphi = \frac{\kappa}{2}\rho_D, \label{eq:PhantomEq}
\end{equation}
where $ \rho_D $ is the charge density of the phantom field $ \varphi $. He introduces $ \rho_D $ as the source of the dark energy. When considering an astrophysical scale of the manifestation of the dark energy, the phantom field's potential can be neglected. Hence, within the derivation of the Einstein-perfect fluid-phantom field system's interior and exterior solutions, one can consider the vanishing potential. It is also considered that there is no interaction between the phantom field and the ordinary matter.

The ansatz of the metric solution is given by
\begin{equation}
ds^2 = -e^{2U}dt^2 + e^{-2U+2\lambda}\left[e^{-2\chi}dr^2 +r^2\left(d\theta^2 +\sin^2\theta d\phi^2 \right) \right],  \label{eq:ansatzmetric}
\end{equation}
where the functions $ U,\lambda, \chi $ depend on coordinate $ r $ only. Moreover, the phantom field $ \varphi $ and charge density $ \rho_D $ are also dependent on radial coordinate only. For the ordinary matter, it is imposed that $ p,\rho $ are dependent on $ r $ only while $ u_\mu dx^\mu = -e^{U}dt $. To produce an exact interior solution of the Einstein-phantom field system (\ref{eq:EinsteinPhantomEq}), one can employ the mathematical methods as given in Ref. \cite{YazadjievModPhysLettA}. We need CDS as a seed metric to generate a new interior solution with an additional phantom field contribution. The resulting interior solution is given as follows
\begin{eqnarray}
ds^2_{int} &=& -e^{2\lambda c} dt^2 + e^{-2\lambda (c-1)} \bigg(\frac{dr^2}{1- \frac{2 C}{R^2}r^2} +r^2 d\theta^2 \nonumber\\
& & + r^2 \sin^2\theta d\phi^2  \bigg), \label{eq:interiorEinsteinPhantom}
\end{eqnarray}
where
\begin{equation}
e^{\lambda} = \frac{3}{2}\left(1- 2C \right)^{\frac{1}{2}}-\frac{1}{2}\left(1- \frac{2C}{R^2}r^2 \right)^{\frac{1}{2}}, 
\end{equation}
\begin{equation}
c= \frac{M}{m},~~ C= \frac{m}{R},~~ m =\sqrt{M^2 -D^2},
\end{equation}
and $ M , D, R $ are the mass, dark charge and the star radius, respectively. Furthermore, the pressure, energy density, charge density and the phantom field are given by \cite{YazadjievPRD2011}

\begin{equation}
 p = \frac{3C}{4\pi R^2}e^{2\lambda (c-1)}\left[\frac{\left(1- \frac{2 C}{R^2}r^2 \right)^{\frac{1}{2}}-\left(1- 2 C \right)^{\frac{1}{2}}}{3\left(1- 2C \right)^{\frac{1}{2}}-\left(1- \frac{2C}{R^2}r^2 \right)^{\frac{1}{2}}} \right], \label{eq:pressure}\
\end{equation}
\begin{equation}
\rho = \frac{3M}{4\pi R^3}e^{2\lambda (c-1)} + 3 (c-1)p, \label{eq:endensity}\
\end{equation}
\begin{equation}
\rho_D = \frac{3D}{4\pi R^3}e^{2\lambda (c-1)} + \frac{3D}{m}p, \label{eq:chardensity}\
\end{equation}
\begin{equation}
 \varphi = \frac{D}{m}\lambda \label{eq:phantom}.\
\end{equation}
\textcolor{black}{DES satisfies the following equilibrium equation
\begin{equation}
\partial_r p + (\rho+p)\partial_r U = \rho_D \partial_r\varphi.
\end{equation}
Above equation explains that the dark energy term provides effective pressure to balance the gravitational force and hydrodynamic force in the left hand side. DES also} satisfies the following Buchdahl inequality \cite{YazadjievPRD2011}
\begin{equation}
C \leq \frac{4}{9},\label{eq:BuchdahlIneq}
\end{equation}
which is similar as CDS. This interior solution will reduce to CDS when the dark charge vanishes. On the surface of the star ($ r=R $), the interior solution matches continuously with the exterior solution which is given by
\begin{eqnarray}
ds^2_{ext} &=& - \left(1- \frac{2m}{r} \right)^{c}dt^2 + \left(1- \frac{2 m}{r} \right)^{-(c-1)} \bigg[\frac{dr^2}{1- \frac{2 m}{r}} \nonumber\\
&&+r^2\left(d\theta^2 +\sin^2\theta d\phi^2 \right) \bigg] . \label{eq:extphantommetric}\
\end{eqnarray}
\textcolor{black}{This is the asymptotically flat phantom black hole solution. When $ c=1 $, it will reduce to the Schwarzschild black hole. Note that this black hole solution corresponds with the phantom field through the charge $ D $ where the phantom field is given by
\begin{equation}
\varphi = \frac{D}{2m}\ln \left(1- \frac{2m}{r} \right). \label{eq:extphantomfield}
\end{equation}
At spatial infinity, the phantom field will vanish.} 

In \cite{YazadjievPRD2011}, Yazadjiev also considers an extremal condition of DES. Nevertheless, this extremality is distinct from the extremality on the black holes, which indicates the condition where the horizons coincide with being one horizon. The extremality on DES refers to $ |D| \rightarrow M $ when keeping $ R $ fixed. In this limit, the space-time metric of DES  reduces to
\begin{eqnarray}
ds^2 &=& -e^{-\frac{M}{R}\left(3-\frac{r^2}{R^2} \right)}dt^2 + e^{\frac{M}{R}\left(3-\frac{r^2}{R^2} \right)} \bigg(dr^2 +r^2 d\theta^2 \nonumber\\
&&+ r^2\sin^2\theta^2 d\phi^2 \bigg). \label{eq:extremalDES}
\end{eqnarray}
This corresponds to the following conditions
\begin{equation}
\rho =\frac{3M}{4\pi R^3}e^{-\frac{M}{R}\left(3-\frac{r^2}{R^2} \right)},~~\rho_D=\pm\rho,~~ p=0.
\end{equation}
The phantom field is then given by
\begin{equation}
\varphi = \mp \frac{M}{2R}\left(3-\frac{r^2}{R^2} \right).
\end{equation}
\textcolor{black}{Note that  $D \rightarrow M$ corresponds with $ \varphi=U $ and $ \rho_D =\rho $ while $D \rightarrow -M$ corresponds with $ \varphi=-U $ and $ \rho_D =-\rho $.} The extremal DES (\ref{eq:extremalDES}) matches continuously with the following exterior phantom black hole, $ |D| \rightarrow M $, 
\begin{equation}
ds^2 = -e^{\frac{2M}{r}}dt^2 + e^{\frac{2M}{r}} \left[dr^2 +r^2\left(d\theta^2 +\sin^2\theta^2 d\phi^2 \right) \right],
\end{equation}
and the phantom field now is
\begin{equation}
\varphi = \mp \frac{M}{r}.
\end{equation}
In this work, we do not intend to discuss furthermore DES's extremal condition. One can see the \textcolor{black}{elaborate} explanation of this extremal DES in Ref. \cite{YazadjievPRD2011}.

\section{Energy Conditions, Stabilities and Redshift}
\label{sec3}
\subsection{Energy conditions}

In the original paper \cite{YazadjievPRD2011}, the author claims that the interior solution in Eq. (\ref{eq:interiorEinsteinPhantom}) represents a compact object containing an ordinary matter and dark energy with the dark charge $ D $ while $ \rho_D $ is the dark source. As we know in cosmology, the phantom field is one of the models to describe dark energy that will produce an accelerating expansion of the universe with the EoS parameter $ \omega  <-1 $ \cite{Copeland2006}. However, to justify an object whether contains a dark energy or not, we need some conditions that really describe the properties of dark energy.

\begin{figure}[!t]
\centering
\includegraphics[scale=0.55]{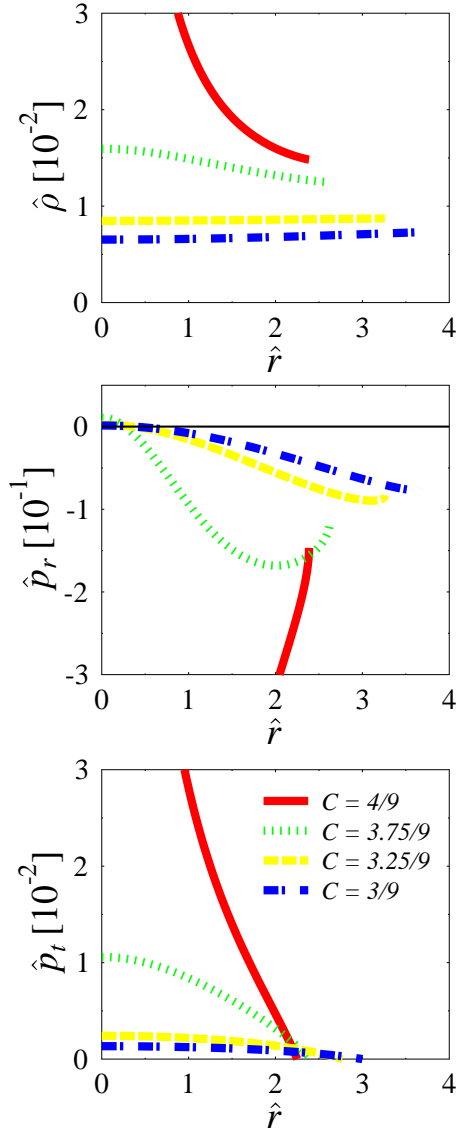}
\caption{Comparison of $\hat{\rho} ~(\rho m^2), \hat{p}_r ~(p_r m^2), $ and $\hat{p}_t ~(p_t m^2)$ in different compactness. We set $ M=5 $ and $ R=9 $. Note that $ \hat{r}=r/m $.}
\label{fig:rhoprpt}
\end{figure}

\begin{figure}[!t]
\centering
\includegraphics[scale=0.55]{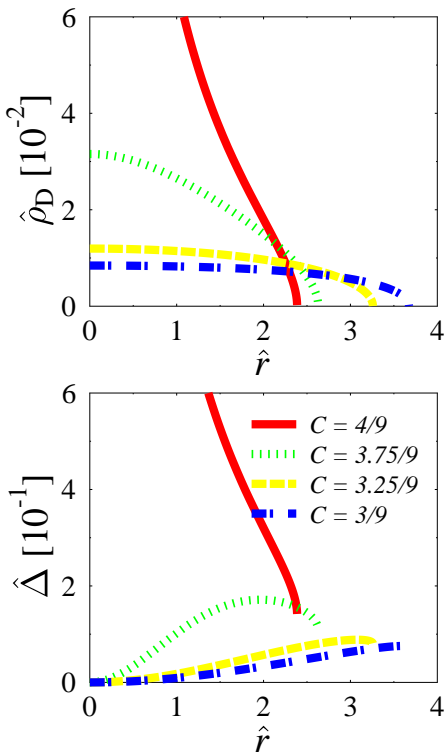}
\caption{Comparison of $\hat{\rho}_D ~(\rho_D m^2)$ and $\hat{\Delta} ~(\Delta m^2)$ in different compactness ($ M=5 $, $ R=9 $).}
\label{fig:rhoDani}
\end{figure}

Regarding the properties of dark energy, one can check the energy condition of \textcolor{black}{the stars}. The energy conditions are crucial to verify the existence of realistic matter distribution in any astrophysical objects. The energy conditions are divided into four, i.e. Null Energy Condition (NEC), Weak Energy Condition (WEC), Dominant Energy Condition (DEC) and Strong Energy Condition (SEC). In mathematical form, these energy conditions are given as follows \cite{BharRahamanRayEPJC2015}
\begin{eqnarray}
\text{NEC} &:& \rho + p_r \geq  0,~\rho + p_t \geq  0, \label{eq:NEC}\\
\text{WEC} &:& \rho + p_r \geq  0,~\rho \geq 0 ,~\rho + p_t \geq  0, \label{eq:WEC}\\
\text{DEC} &:& \rho \geq |p_r|,~\rho \geq |p_t| , \label{eq:DEC}\\
\text{SEC} &:& \rho + p_r \geq  0,~\rho + p_r + 2p_t \geq  0, \label{eq:SEC}\
\end{eqnarray}
where $ p_r$ and $p_t $ are the pressure in radial and tangential direction, respectively. WEC stipulates that the energy density is required to be non-negative measured by the corresponding observer. DEC stipulates that the pressures do not exceed the energy density, so that the sound speed in the fluid is always less than the speed of light. However, SEC may not be satisfied that denotes the strong repulsion of gravity. We can investigate the energy conditions for this DES of which if this star contains dark energy, it shall violate SEC (\ref{eq:SEC}). For DES, note that the radial and tangential pressures are given by
\begin{equation}
p_r = p - 2g^{rr}(\partial_r \varphi)^2 ,~~~ p_t = p, \label{eq:allpressure}
\end{equation}
respectively. Furthermore, as we can see, DES is an anistropic star where the anisotropy factor is given as follows
\begin{equation}
\Delta =  2g^{rr}(\partial_r \varphi)^2 =\frac{2D^2 r^2}{R^6}e^{2\lambda \left(c-2 \right)}. \label{eq:anisotropy} \
\end{equation}
This anisotropy contributes only to the radial pressure.

We can see the comparison of $ \rho$, $p_r$ , and $p_t $ on Fig. \ref{fig:rhoprpt} and $ \rho_D$ and $\Delta $ on Fig. \ref{fig:rhoDani} for several values of $ C $ when we set $ M=5 $ and $ R=9 $. Note that this star's pressures are not linear with the energy density as the common model for dark energy in our universe. For $ C =4/9 $, the pressures and energy density are singular at the center of the star.
The enormous pressure in the center could cause a change on the topology as noted in \cite{LoboCQG2006} in which this may produce a tunnel and transform DES into a wormhole \cite{MorrisThorne1988,VisserAIP1995} as we can see that the phantom energy may aid the traversable wormholes \cite{SushkovPRD2005,LoboPRD2005}. For $ C=3.25/9 $ and $ C=3/9 $, the center's energy density is lower than on the surface. The radial pressure is minimal at the center and becomes more negative on the surface, except for $ C=4/9 $ while the tangential pressure is monotonically decreasing for all $ C $.  Similarly with the tangential pressure, the charge density is also decreasing.

Then we can see that the anisotropy is higher near the surface than the center for the compactness below $ 4/9 $. Since the anisotropy is positive, it is clear that the anisotropic force is repulsive rather than attractive. The phantom field's contribution is not dominant near the star's center for $ C $ below the Buchdahl limit.

\begin{figure}[!t]
\centering
\includegraphics[scale=0.55]{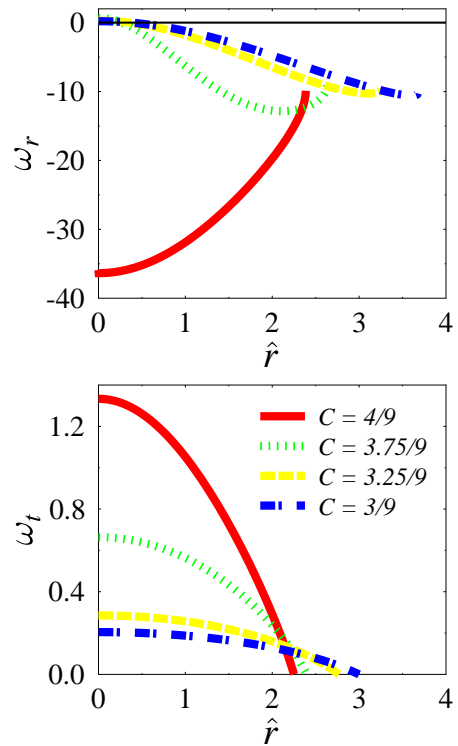}
\caption{EoS parameter for radial ($\omega_r $) and tangential ($\omega_t $) direction in different compactness ($ M=5 $, $ R=9 $).}
\label{fig:eos}
\end{figure}

For the present universe where the dark energy supports its accelerated expansion, EoS is given as $\omega =  p /\rho $ where the parameter $ \omega $ needs to be negative in the range $ -1 \leq \omega \leq -1/3 $. However, in the current observation, the EoS parameter can still be lower than $-1$. It is a condition for a phantom scalar field where the kinetic energy term has a negative sign. For DES, the EoS parameter is dependent on radial coordinate, not constant. Because of the anisotropic pressures, it is convenient to write the equations of state as follows
\begin{equation}
\omega_r = \frac{p_r}{\rho}, ~~~ \omega_t = \frac{p_t}{\rho}. \label{eq:eosds}
\end{equation}
We denote $ \omega_r $ as radial EoS parameter and $ \omega_t $ as tangential EoS parameter of DES. We show its profile in Fig. \ref{fig:eos}.
From those panels, we can see that the condition $ \omega_r <-1 $ is not satisfied near the center of DES for $ C $ smaller than Buchdahl limit. Moreover, $ \omega_t $ is always positive.

We have mentioned beforehand that one property of the existence of dark energy in an astrophysical object is it should violate the SEC \cite{ChanSilvaMPLA2009}. Hence, we need to investigate its energy conditions. 
\begin{figure*}[!t]
\centering
\includegraphics[scale=0.55]{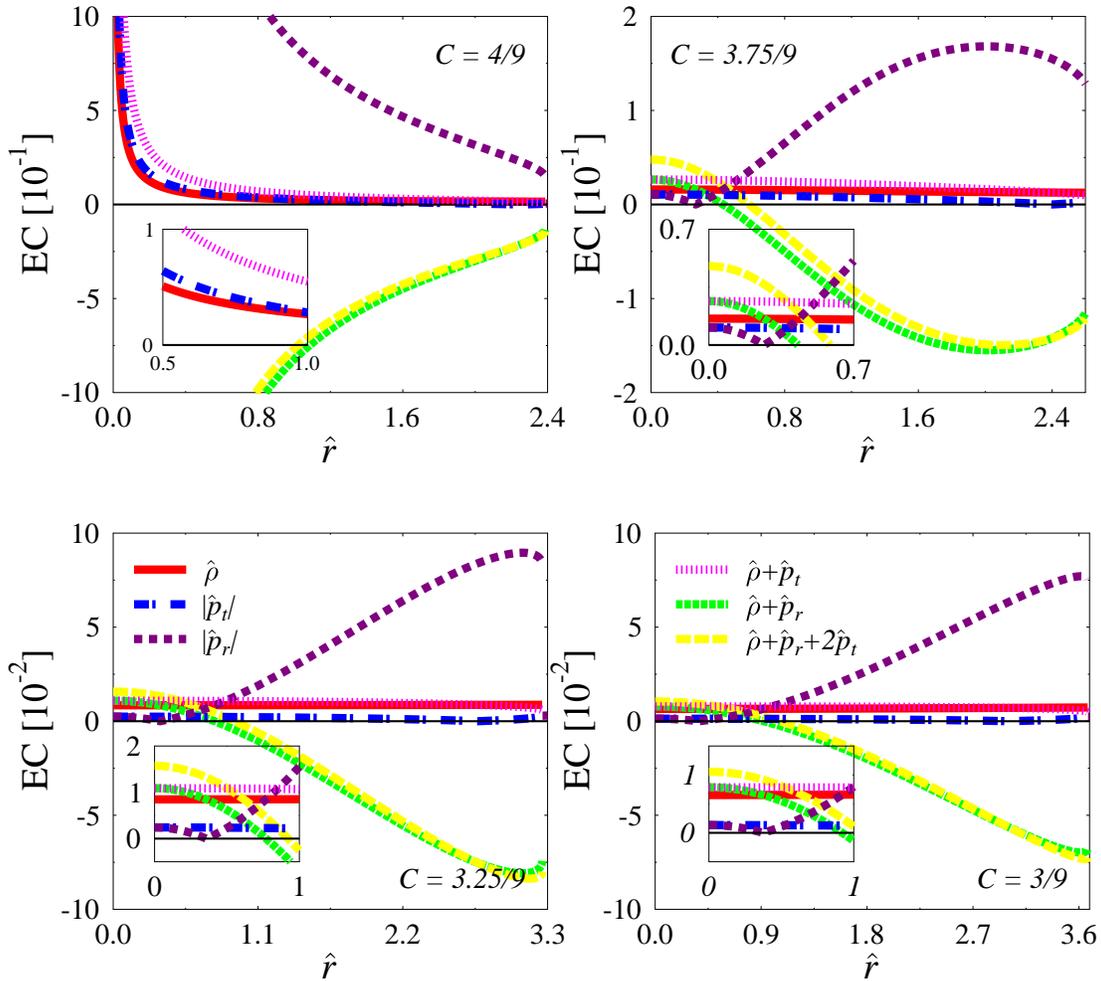}
\caption{Comparison of the energy conditions (EC) in different compactness ($ M=5 $, $ R=9 $).}
\label{fig:encon}
\end{figure*}
As we can see from Fig. \ref{fig:encon}, DEC is not satisfied. Furthermore, the condition $ \rho \geq 0 $ is satisfied. Regarding SEC, it is fully satisfied for $ C = 4/9 $ for dark energy condition. For compactness less than $ C =4/9 $, it is obviously seen that near the center of the star, SEC is not fully violated. Hence, we may say that DES still contains dark energy but it is not fully distributed in the interior of the star for $ C $ below the Buchdahl limit.

\subsection{Causal conditions}
\begin{figure}[!t]
\centering
\includegraphics[scale=0.55]{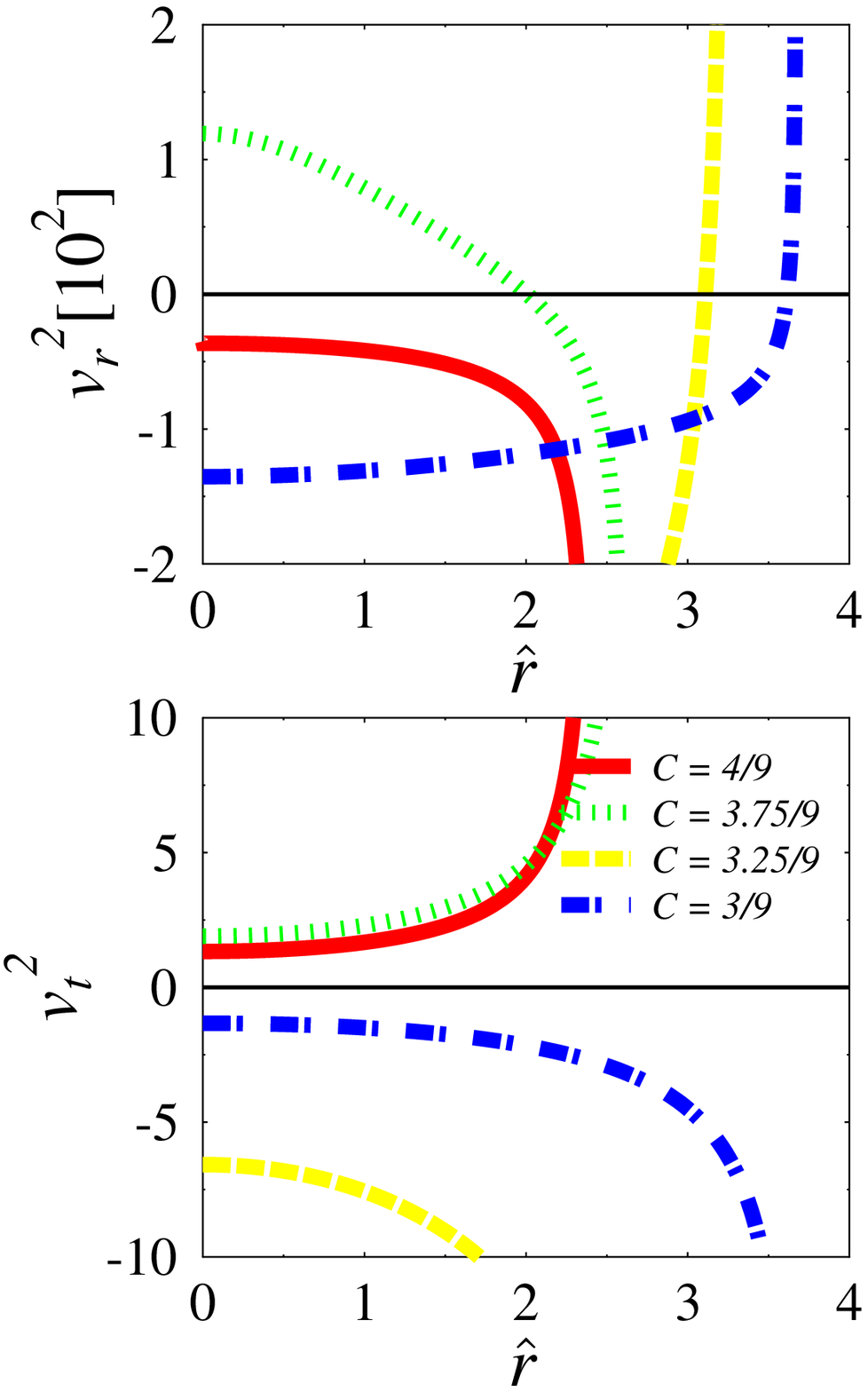}
\caption{Causal conditions in different compactness ($ M=5 $, $ R=9 $). }
\label{fig:causal}
\end{figure}

According to the initial observer, the relativistic principle of causality says that the cause has to precede its effect. So, the cause and its consequence are separated by a time-like interval. If a time-like interval separates two events, it means that a signal could be sent between them at less than the speed of light. If there is a signal that moves faster than the speed of light, it will violate the causality. So, special relativity will not allow an object to have a speed faster than light. For an anisotropic structure, the causal conditions are given as follows
\begin{equation}
0 \leq v_r^2 = \frac{dp_r}{d\rho} \leq 1, ~~~0 \leq  v_t^2 = \frac{dp_t}{d\rho} \leq 1. \label{eq:causality}
\end{equation}
The plots of causal conditions of DES are shown in Fig. \ref{fig:causal} where the top panel is for the radial speed, and the bottom panel is for the tangential speed. It is seen that the causal conditions are violated.   

We know that CDS also does not satisfy the causal conditions because of the incompressible energy density. However, CDS can be a representative toy model for compact objects that respect the Buchdahl limit and have a chance to become the ultra-compact objects but still violate the causal conditions \cite{PaniFerrariCQG2018}. Because the Buchdahl limit's derivation does not use the causal conditions, it is plausible that a compact object may not satisfy one of those conditions. We find that DES does not satisfy the causal conditions but still respects the Buchdahl limit. Hence, we may say that DES may also be a good and representative toy model to study an ultra-compact object containing ordinary matter and dark energy.

\subsection{Stabilities due to convective motion and gravitational cracking}

\begin{figure}[!t]
\centering
\includegraphics[scale=0.55]{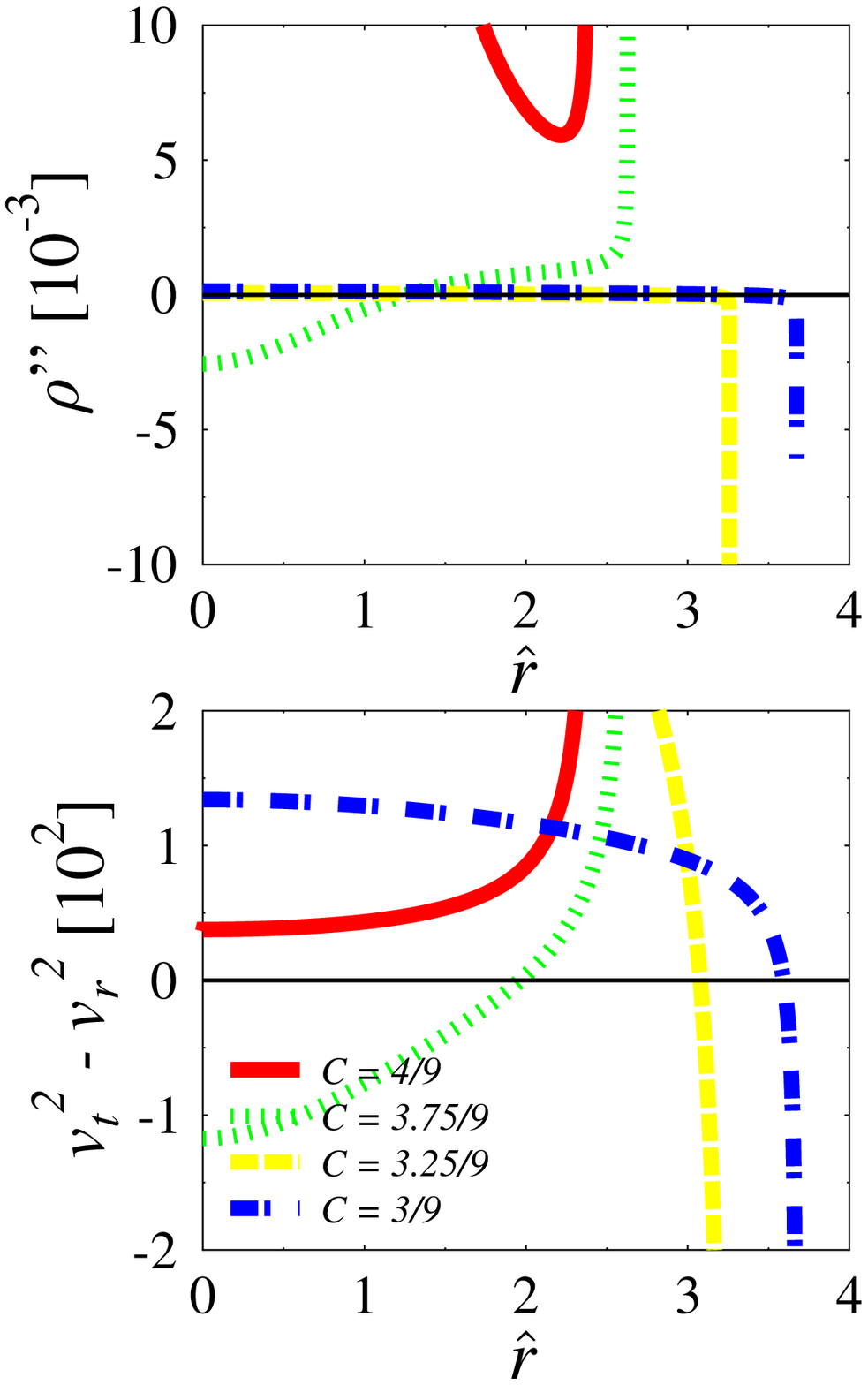}
\caption{Stability conditions due to convective motion (top) and due to gravitaional cracking (bottom) in different compactness ($ M=5 $, $ R=9 $).}
\label{fig:stability}
\end{figure}

We also consider the stabilities of DES due to the convective motion, and gravitational cracking \cite{HectorEPJC2018}. The stability due to the convective motion implies that when a fluid element is displaced downward, the fluid element will sink, and the star will be unstable. If the fluid element's density is less than its surroundings, it will float back, so that the star will be stable to the convective motion. The criterion of stability due to the convective motion is given by
\begin{equation}
\rho''=\frac{d^2\rho}{dr^2}\leq 0. \label{eq:convective}
\end{equation}
Hence, when the fluid of DES satisfies this circumstance, the star is stable due to the convective motion. This criterion for DES is shown in Fig. \ref{fig:stability} (top panel). It can be seen that $ \rho'' $ is almost approaching zero, except for $ C =4/9 $. Hence, for $ C $ below the Buchdahl limit, the fluid is almost stable due to the convective motion. We know that $ \rho'' =0 $ because the energy density is constant for CDS. So, as a toy model, CDS can be stable due to the convective motion. However, the phantom scalar field's presence does not significantly impact the star because it omits the stability due to the convective motion compared to CDS.

Besides the convective motion, as we have mentioned before, we also consider the stability due to the gravitational cracking. The anisotropic matter distribution could cause the gravitational cracking instability to the stars \cite{HectorEPJC2018} in which this instability does not occur on CDS. The cracking instability determines the tidal acceleration profiles produced by the perturbations of the energy density and the anisotropic pressures identifying the sign of the total force in the system. The condition to be satisfied by a fluid distribution to avoid gravitational cracking is given by
\begin{equation}
-1 \leq v_t^2 - v_r^2 \leq 0. \label{eq:gravcrack}
\end{equation}
For DES, the plot of the gravitational cracking is shown in Fig. \ref{fig:stability} (bottom panel).
From that figure, we may conclude that DES is not stable due to the gravitational cracking.

\subsection{Redshift}
\begin{figure}[!t]
\centering
\includegraphics[scale=0.55]{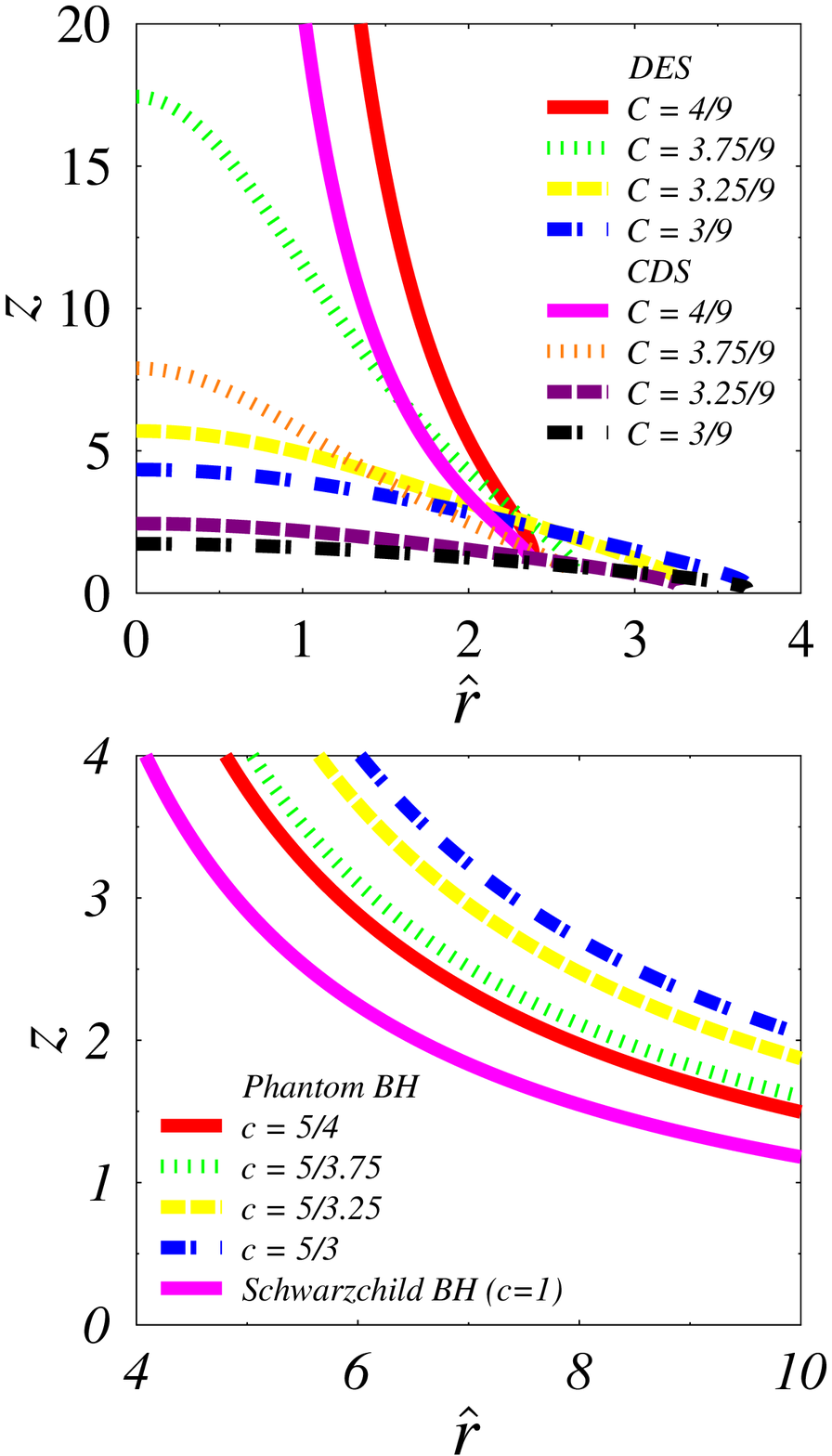}
\caption{Redshift for interior (top) and exterior (bottom) solutions  in different compactness ($ M=5 $, $ R=9 $).}
\label{fig:redshift}
\end{figure}

The redshift of DES and its exterior solution can be computed using
\begin{equation}
z = \frac{1}{(-g_{tt})^{1/2}} -1 . \label{eq:redshift}
\end{equation}
The plot of the redshift for different value of compactness is given in Fig. \ref{fig:redshift} where the top panel is for CDS and DES, and the bottom panel is for the exterior solutions. Those figures show that the redshift is monotonically decreasing to the increasing of the radius of the star. The values of surface redshift $z_s$ of DES and CDS for each compactness are given in Table \ref{tab:table1}.

\begin{table}[h!]
	\centering
    \caption{Surface redshift $z_s$ for DES and CDS.}
    \label{tab:table1}
    \begin{tabular}{c c c c c c c} 
    \hline
    \hline
 ~~~~~~~~  &  &~~~~~~~~~~~~~~~& $z_s$ &~~~~~~~~~~~~~~~ & $z_s$ &~~~~~~~~ \\
     & \textbf{$C$} &  & CDS & & DES & \\
      \hline
     & 4/9 & & 1.0142 & & 1.3995 & \\
      
    &  3.75/9 & &  0.6412 & &  0.9359 &\\
      
    &  3.25/9 & & 0.2658 & & 0.4371 &\\
      
    &  3/9 & & 0.1581 & & 0.2771 &\\
      \hline
      \hline
    \end{tabular}
\end{table}
From Table \ref{tab:table1}, it is seen that the surface redshift for DES is larger than CDS for the same compactness. Hence, the phantom field's presence prolongs the wavelength that is measured by the observer at infinity. This finding is an excellent sign to observe the existence of such dark star.

\section{Echoes from Ultra-compact Dark Energy Star}
\label{sec4}
\subsection{Echo time and echo frequency}
The production of gravitational waves' echoes requires a photon sphere, which implies that the object's compactness must be greater than $ 1/3 $. For the compactness greater than $ 1/3 $, we note this object as an ultra-compact object, but it still satisfies the Buchdahl inequality Eq. (\ref{eq:BuchdahlIneq}). In investigating gravitational echoes, it is important to calculate the echo frequency. CDS is a good toy model to study the echo frequency as it has been calculated in \cite{PaniFerrariCQG2018,UrbanoVeermaeJCAP2019} where its frequency is highest compared to other fluid stars with the same mass and radius. The echo frequency can be roughly estimated from the inverse of the time $ \tau $ for a massless test particle traveling from the unstable light ring or photon sphere to the center of the star. The echo frequency is $ f_{echo} = \pi/\tau_{echo} $ while the echo time is given by \cite{PaniFerrariCQG2018,UrbanoVeermaeJCAP2019,MannarelliPRD2018}
\begin{equation}
\tau_{echo} = \int^{3m}_{0} \left(-\frac{g_{rr}(r)}{g_{tt}(r)} \right)^{1/2} dr. 
\end{equation}
For CDS, we can find \cite{UrbanoVeermaeJCAP2019}
\begin{eqnarray}
\frac{\tau_{echo}}{m} &=&  \frac{\cot^{-1}{\left(\sqrt{\frac{4}{C}-9} \right)} + \tan^{-1}{\left(3/\sqrt{\frac{4}{C}-9} \right)}}{C^2 \sqrt{\frac{4}{C}-9}} \nonumber\\
&&- 2\ln \left(\frac{1}{C}-2 \right)+3- \frac{1}{C}.
\end{eqnarray}
It is worth noting that $ m=M $ for CDS. It is easy to check that when $ C \rightarrow 4/9 $, $ \tau_{echo} \rightarrow \infty $.

DES with zero dark charge will become CDS. Hence, it is interesting to examine the dark charge's impact on the echo frequency of DES. So, we will compare our result with the result of CDS. It is worth noting that the unstable light ring of DES is different from CDS as we derive in Appendix \ref{app:photonsphere}. For DES, the echo time for a massless test particle from the photon sphere to the center of the star can be calculated as
\begin{eqnarray}
\tau_{echo} &=& \int^R_0 \frac{\left[\frac{3}{2}\sqrt{1-2C}-\frac{1}{2}\sqrt{1-\frac{2Cr^2}{R^2}} \right]^{1-2c}}{\sqrt{1-\frac{2Cr^2}{R^2}}} dr \nonumber\\
&&+ \int^{M(2c+1)}_R \left(1-\frac{2m}{r} \right)^{-c} dr,
\end{eqnarray}
where the photon sphere is located on  $r = M(2c+1) $ (Appendix \ref{app:photonsphere}). The integration of the interior solution above cannot be solved analytically. Nonetheless, we can solve the exterior part analytically with the hyperbolic function. By taking $ r = m \hat{r} $ and integrating the exterior part, we can write the time as
\begin{eqnarray}
\frac{\tau_{echo}}{m} &=& {\LARGE \int}^{R}_0 \frac{\left[\frac{3}{2}\sqrt{1-2C}-\frac{1}{2}\sqrt{1-2C^3\hat{r}^2 } \right]^{1-2c}}{\sqrt{1-2C^3\hat{r}^2}} d\hat{r} \nonumber\\
&- & \frac{(-2C)^{-c}}{C(1+c)}F_a + \frac{(-2)^{-c} \left[c(2c+1) \right]^{c+1}}{1+c}F_b, \
\end{eqnarray}
where
\begin{eqnarray}
F_a &=& {_2F_1}\left(c,1+c;2+c;\frac{1}{2C} \right),\nonumber\\
F_b &=& {_2F_1}\left(c,1+c;2+c;\frac{(2c+1)c}{2}\right).\nonumber\
\end{eqnarray}
and ${_2F_1}(a,b;c;d)$ is a hypergeometric function. It is obviously seen that $ \tau_{echo} $ is dependent on $ C $ and $ c $ unlike CDS that is dependent on $ C $ only. The integration of the echo time for exterior part yields a complex function.
\begin{figure*}[!t]
\centering
\includegraphics[scale=0.55]{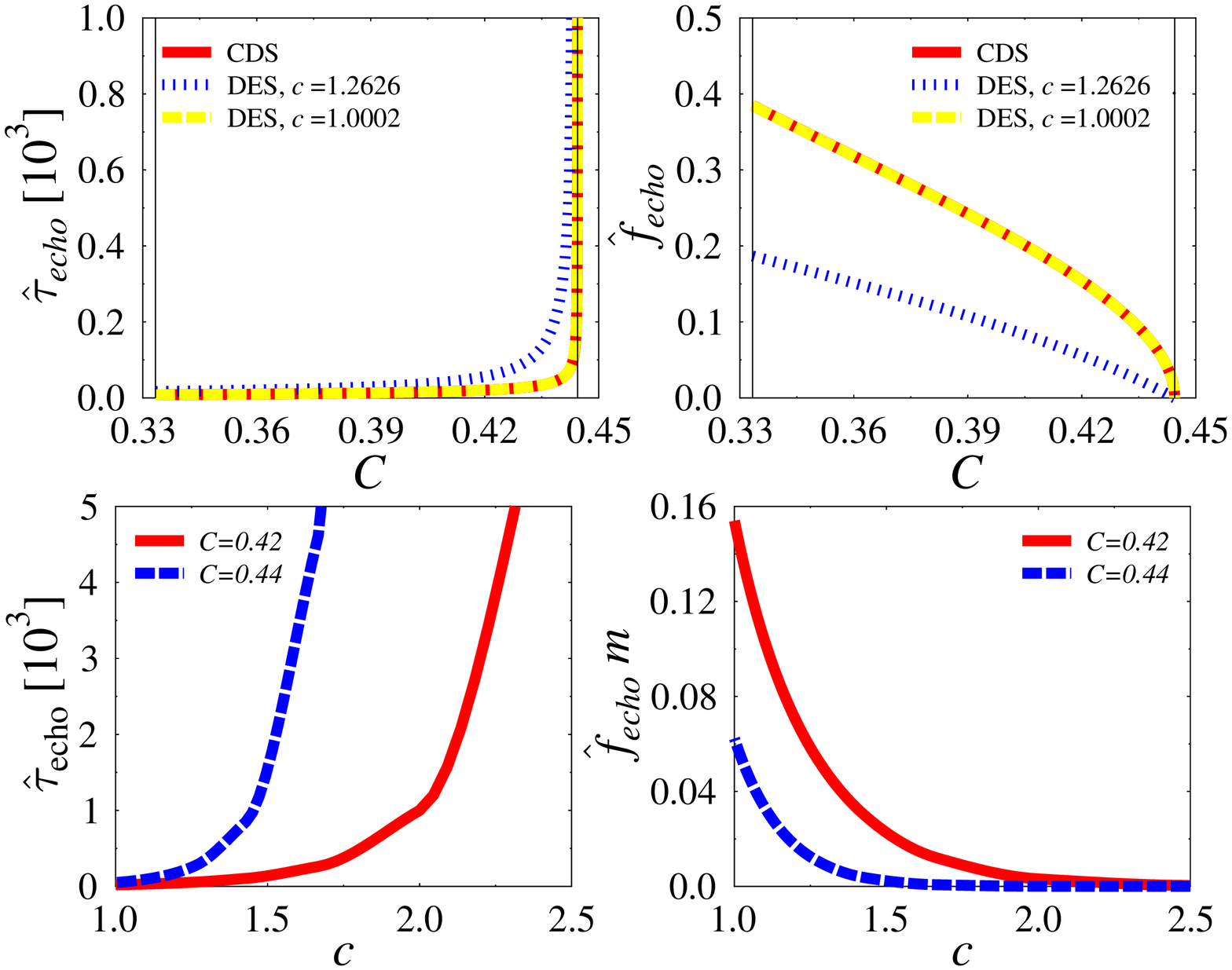}
\caption{Echo time ($ \hat{\tau}_{echo} =\tau_{echo}/m$) and echo frequency as a function of $ C $ (top) and as a function of $ c $ (bottom).}
\label{fig:taufreq}
\end{figure*}

Unlike the echo time of CDS, due to the phantom field's presence as the source of dark energy, there is an imaginary contribution to the echo time. However, the imaginary contributions coming from the integration from the surface to the photon sphere cancel each other. We show the phantom field's impact by comparing it with CDS shown in the above panels of Fig. \ref{fig:taufreq}. We show the echo time and frequency as a function of $ C $ with $ c=1.2626 $ and $ c=1.0002 $. Both echo times go to infinity when approaching the Buchdahl limit. When $ c>1 $, the echo time is bigger than CDS while the frequency is inversely proportional. It denotes that the presence of a phantom field as the dark energy model delays the gravitational echoes. The larger $ c $ (larger $ D $) delays the echoes more. When $ c \approx 1 $, the echo time of DES approaches CDS' echo time. In the bottom panels of Fig. \ref{fig:taufreq}, we show the echo time and echo frequency as a function of $ c $ for DES when $ C=0.42 $ and $ C=0.44 $. From the bottom panels of Fig. \ref{fig:taufreq}, it can be seen that the echo time becomes larger for larger $ c $. On the other hand, the larger $ c $ will expedite the echoes. 

\subsection{Effective potential}

CDS enjoys the presence of a potential well in between the light ring at $ r=3M $ and the center of the star $ r=0 $ \cite{PaniFerrariCQG2018,UrbanoVeermaeJCAP2019}. It means that CDS has a second light ring at the minimum of the potential. In the previous subsection, we have shown the echo time and echo frequency for both CDS and DES. We could infer that there is a similar feature regarding the gravitational echoes on both star models. Hence, DES must enjoy a potential well as CDS, yet with different conditions due to the phantom field's existence. 

We focus on the axial perturbation of DES (\ref{eq:interiorEinsteinPhantom}) and its exterior solution (\ref{eq:extphantommetric}). For a spherically symmetric space-time (\ref{eq:ansatzmetric}), we may apply the wave equation as shown in Appendix {\ref{app:perturbation}. To use  wave equation (\ref{eq:waveequation}) for space-time metric (\ref{eq:ansatzmetric}), we need to transform
\begin{equation}
v(r) \rightarrow 2U(r), ~~~ \xi(r) \rightarrow -2U(r)+2\lambda(r)-2\chi(r). 
\end{equation}.
For DES, we know that $ P(r) \rightarrow p_r(r) $. Hence, we can find the effective potential as
\begin{eqnarray}
V_{s,l}(r) &=& e^{2U(r)}\left\{\frac{l(l+1)}{r^2}+\frac{1-s^2}{re^{-2U(r)+2\lambda(r)-2\chi(r)}}\right. \nonumber\\
& &\times  [\chi'(r)-\lambda'(r)] -4\pi [p_r(r) -\rho(r)]    \bigg\} .\label{eq:effectivepotds}
\end{eqnarray}
The tortoise coordinate is now $ dr_* = e^{\lambda(r)-\chi(r)-2U(r)}dr $. Using Eq. (\ref{eq:effectivepotds}), we can analyze the effective potential due to the presence of the phantom field.

To analyze the effective potential, we use the ordinary dimensionless radial coordinate $ \hat{r} $ and dimensionless tortoise coordinate $ \hat{r}_* $. Firstly, the comparison of the effective potential for CDS and DES is explicitly given in the top panel of Fig. \ref{fig:VeffSchvsDs} for $C=0.44$ and $ c=1.2626 $. Note that the discontinuity on the effective potential denotes the surface of the stars. Besides, for Schwarzschild black hole and the phantom black hole, the effective potentials are shown in the bottom panel of Fig. \ref{fig:VeffSchvsDs}. It is seen that the effective potentials of the stars form a potential well between the center of the star to the unstable circular orbit or light ring. Both potentials enjoy the presence of a potential well. This fact means that DES also possesses a second (stable) the light ring like CDS \cite{UrbanoVeermaeJCAP2019}. From this fact, we may observe that DES possesses an unstable light ring, one of the conditions to produce the gravitational echoes. It also possesses a stable light ring at the minimum of the potential well. The difference between CDS and DES is the peak of the potential for the same compactness. The potential well of DES is higher (deeper) than CDS, which corresponds to the echo time. The echo time for DES is longer than CDS because of this. It means that the gravitational waves need more time to travel in the potential well of DES than in the potential well of CDS.
\begin{figure}[!t]
\centering
\includegraphics[scale=0.56]{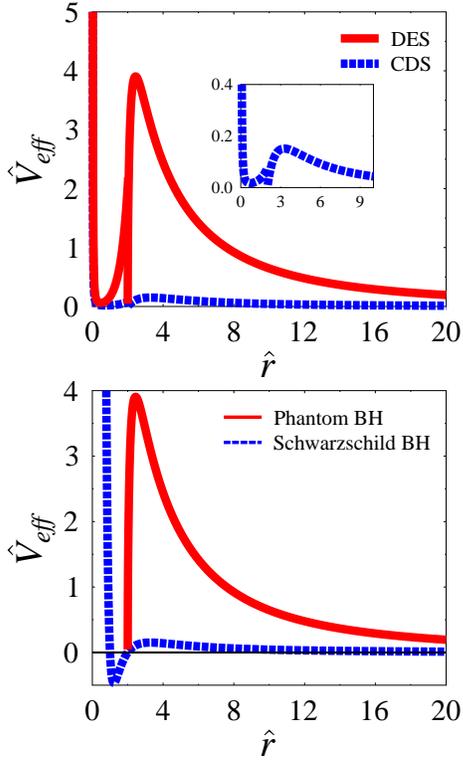}
\caption{Effective potential ($ \hat{V}_{eff} = V_{eff} m^2 $) of CDS and DES for $ C=0.44$ and $ c=1.2626 $.}
\label{fig:VeffSchvsDs}
\end{figure}
\begin{figure}[!t]
\centering
\includegraphics[scale=0.56]{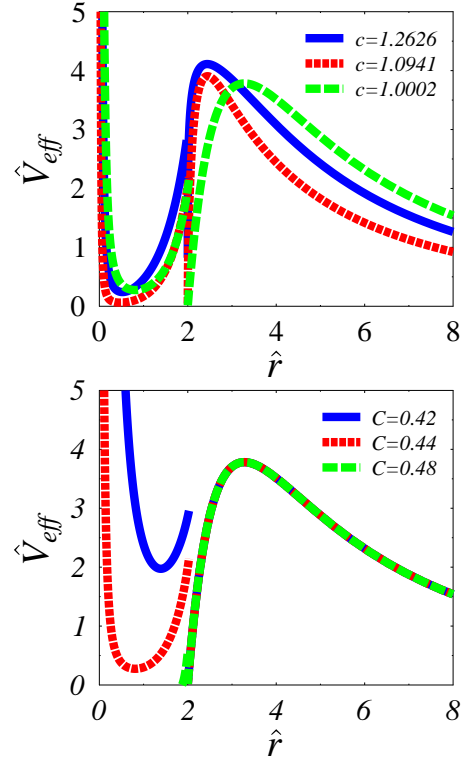}
\caption{Effective potential of DES for different $ c $ and $ C $.}
\label{fig:VeffDs}
\end{figure}

In Fig. \ref{fig:VeffDs}, we show the effective potential for different $ c $ when $ C=0.44 $ in the top panel and for different $ C $ for $ c=1.0002 $ in the bottom panel for DES. The top panel shows that the peak of the potential decreases and the width of the potential well increases, but it is not significant. Then for the bottom panel, we can infer that the depth of potential well increases as the compactness increases. When the compactness approaches the black hole's compactness $ (C=0.5) $, the effective potential resembles the black hole's effective potential. The gravitational waves take more time to travel in the potential well with a higher $ c $ because the potential is relatively wide. While in higher $ C $, the gravitational waves take more time to travel since the potential well is deep.

For lative Schwarzschild exterior solution, the tortoise coordinate is given by
\begin{equation}
  r_* = r + 2m \ln \left(\frac{r}{2m}-1 \right),
  \label{aba}
\end{equation}
The inverse of Eq. (\ref{aba}) that or $ r(r_*) $ can be obtained using the Lambert $W$ function. The effective potential of CDS in tortoise coordinate is shown in Fig. \ref{fig:VeffCDS}. The difference between CDS and Schwarzschild black holes' effective potentials is that CDS' potential tends to infinity at the center that corresponds with the centrifugal term $ l(l+1)/r^2 $ while the potential of the black hole vanishes at the horizon.
\begin{figure}[!t]
\centering
\includegraphics[scale=0.56]{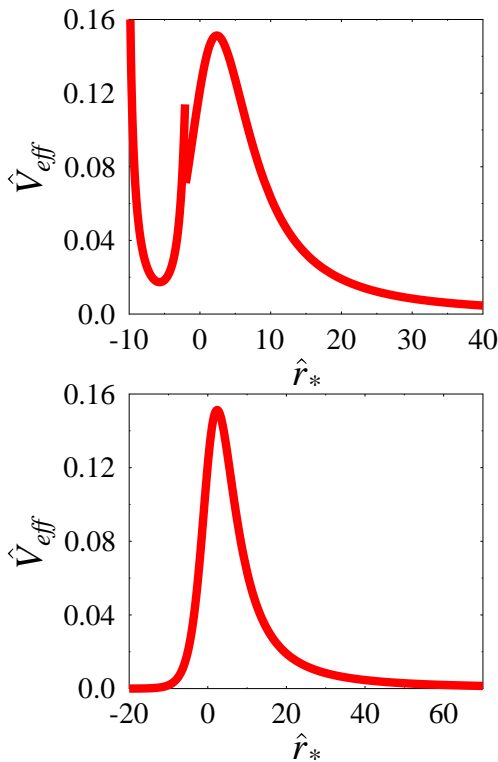}
\caption{Effective potential of CDS in tortoise coordinate for $ C=0.44 $.}
\label{fig:VeffCDS}
\end{figure}
Nonetheless, for the phantom black hole, the tortoise coordinate is explicitly given by
\begin{equation}
r_* = \left(-\frac{r}{2m} \right)^{c} \frac{r}{1+c} {_2F_1\left(c,1+c;2+c;\frac{r}{2m} \right)}, \label{eq:tortoisephantom}
\end{equation}
which is a complex function, and we cannot find the inverse analytically. Hence, the effective potential of DES cannot be described in tortoise coordinate. However, since the effective potential in coordinate $ r $ of DES possesses similar properties with CDS, it is presumable that the potential of DES possesses similar properties. Furthermore, one can also employ the P{\"o}schl-Teller potential to delve the wavefront of gravitational echoes \cite{TestaPani2018PRD} of which this is left for further study.

\section{Gravastar-like Condition of Dark Energy Star}
\label{sec6}
There is a unique and fascinating condition for CDS. It could describe a simple gravitational vacuum condensate star or famously called by gravastar \cite{MazurMottolaPNAS2004,MazurMottolaCQG2015}. As we know, black holes are believed to be the final stage of the gravitational collapse. However, it possesses an event horizon where the physics that we know no longer exists. Gravastar comes as a serious alternative to a gravitational collapse system in which it has no singularity within it. This star brings new interest in the gravitational system. Gravastar is introduced by Mazur and Mottola \cite{MazurMottolaPNAS2004} as a dark, cold, and compact object with a de Sitter condensate phase covered up by a thin shell. The thin shell consists of an ultra-relativistic matter directly adjacent with a total vacuum spacetime or Schwarzschild spacetime. However, it is presented in Ref. \cite{CattoenFaberCQG2005} that gravastar may not have a thin shell when it possesses an anisotropic structure. Gravastar is thermodynamically stable with vanishing entropy in the interior region.

Specifically, gravastar possesses three different segments with three different equations of state,
\begin{eqnarray}
\text{I}.&~& \text{Interior}~~ 0\leq r <R_0, ~~~\rho = -p, \nonumber\\
\text{II}. &~&\text{Shell}~~~~ R_0 < r <R, ~~~ \rho = +p, \\
\text{III}. &~& \text{Exterior}~~~ R< r, ~~~~~~ \rho = p=0. \nonumber\
\end{eqnarray}
The first segment is the interior from the center $ r=0 $ to the radius of $ R_0 $ with de Sitter phase.  In this region, the radius satisfies
\begin{equation}
r< R = R_s = 2m, 
\end{equation}
where $ R_s= 2m $ is the Schwarzschild radius. It is seen that in this circumstance, Buchdahl's inequality is violated. On the other hand, this denotes that the ultra-compact object is formed. The third segment is the vacuum Schwarzschild solution, which is asymptotically flat. To find the second segment, but we need to assume that $ g^{rr}\ll 1 $ at first to gain $ \rho = +p $. We refer to read the original paper \cite{MazurMottolaPNAS2004} about the thin shell's derivation.

In this section, we are going to investigate the similar condition of gravastar to DES. Following \cite{CattoenFaberCQG2005}, due to the anisotropic structure of DES, we assume that the thin shell does not exist for this gravastar-like solution. For DES, the horizon radius is located at
\begin{equation}
R_s = 2C. \label{eq:horizonradiusdarkstar}
\end{equation}
For $ R_s=R=2m $, it will violate the Buchdahl inequality for DES (\ref{eq:BuchdahlIneq}). The compactness $ C $ will be equal to one. This condition corresponds to the following space-time metric
\begin{eqnarray}
ds^2 &=& -\left[\frac{1}{4}\left(1-\frac{r^2}{R^2} \right) \right]^{c} dt^2 + \frac{1}{4}\left[\frac{1}{4}\left(1-\frac{r^2}{R^2} \right) \right]^{- c}dr^2 \nonumber\\
&& + \left[\frac{1}{4}\left(1-\frac{r^2}{R^2} \right) \right]^{1-c}r^2 (d\theta^2 + \sin^2\theta d\phi^2) \label{eq:AdSlikemetric}.\
\end{eqnarray}
If we set $ c=1 $, metric (\ref{eq:AdSlikemetric}) will reduce to \textcolor{black}{dS} metric as the interior of the gravastar. No singularity is present in this space-time metric. It would be fascinating to delve into the symmetry in (\ref{eq:AdSlikemetric}) that we leave for the future works. Furthermore,  the negative pressure will be produced as given by
\begin{equation}
p_t = p  =-\frac{3C}{4\pi R^2}\left[\frac{1}{4}\left(1-\frac{r^2}{R^2} \right) \right]^{c-1}. 
\end{equation}
Hence, now $ p $ is already negative without the anisotropic factor. Now, the anisotropy is given as follows
\begin{equation}
\Delta =\frac{2D^2 r^2}{R^6}\left[\frac{1}{4}\left(1-\frac{r^2}{R^2} \right) \right]^{c-2}.
\end{equation}
Since radial pressure $ p_r = p_t-\Delta $ is still also negative, it is evident that all pressures are negative and depend on coordinate $ r $ in the star's interior. This kind of gravastar is, however, quite distinct from the gravastar constructed in \cite{MazurMottolaPNAS2004}. The energy density is proportional to the tangential pressure where it is given by
\begin{equation}
\rho  =\frac{3(3m-2M)}{4\pi R^3}\left[\frac{1}{4}\left(1-\frac{r^2}{R^2} \right) \right]^{c-1}, 
\end{equation}
as well as with the charge density of DES
\begin{equation}
\rho_D  =-\frac{3D}{2\pi R^3}\left[\frac{1}{4}\left(1-\frac{r^2}{R^2} \right) \right]^{c-1}. 
\end{equation}
We can see obviously that there is no singularity on the pressures, energy density and charge density. Indeed, when $ c=1 $, all pressures and energy density reduce to the ordinary gravastar. EoS parameters of this star are now given as follows
\begin{eqnarray}
&&\omega_r = - \left[\frac{m}{3m-2M}-\frac{32\pi D^2}{3(3m-2M)R}\left(1-\frac{R^2}{r^2}\right)^{-1} \right], \nonumber\\
&& \omega_t = -\frac{m}{3m-2M} .
\end{eqnarray}

It is also intriguing that, unlike the ordinary gravastar, EoS parameters is varied, dependent on mass $ M $, dark charge $ D $ and radius $ R $. In this case, we can find $ \omega_{r,t}< -1 $ for arbitrary $ M,D,R $ which represents phantom dark energy. Moreover, the radial EoS is coordinate-dependent. Hence, this is the condition of the interior of dark gravastar.

The unique solution in the exterior region with the phantom field that approaches flat space-time as $ r \rightarrow \infty $ is a metric that is given  in Eq. (\ref{eq:extphantommetric}) and the phantom scalar field is given in Eq. (\ref{eq:extphantomfield}). It occurs for the radius $ R < r $. In Ref. \cite{MazurMottolaPNAS2004}, the thermodynamic analysis of gravastar is also investigated where the entropy in the interior remains zero using the assumption of vanishing chemical potential. For our dark gravastar, it is obviously seen that $ p+\rho \neq 0 $. However, there must be a dark charge contribution to the thermodynamic equation that shall yield zero entropy in the interior region. However, we leave this analysis for future studies.

\section{Conclusions}
\label{conc}

A class of interior solutions of the Einstein field equation with perfect fluid matter and phantom scalar field has been presented that we call as a dark energy star (DES). DES has an anisotropic structure with negative radial pressure. The anisotropy comes from the phantom field contribution. We have checked the existence of the dark energy represented by the phantom field using energy conditions analysis. We know that the existence of dark energy should violate SEC. We have found that SEC is not thoroughly violated in all ranges of compactness, except in the Buchdahl limit or $ C=4/9 $. It could be seen that for $ C<4/9 $, the phantom field does not fill the center of the star. Hence, the dark energy modeled by the phantom field in DES does not fill up a whole star. However, in principle, the enormous pressure in the center of the star at Buchdahl limit may lead to the production of a tunnel representing a formation of the traversable wormhole \cite{LoboCQG2006}.
 
We also have considered the causal conditions and stabilities of DES due to the convective motion and gravitational cracking. \textcolor{black}{Similar to} CDS, DES with phantom dark energy does not satisfy the radial and tangential speeds' causal conditions. Furthermore, DES is not stable due to the convective motion and the gravitational cracking unlike CDS. The surface redshift has also been investigated. We compared for both surface redshift predicted by CDS and DES. The results show that the surface redshift for DES is larger than that of CDS. It means that the presence of the phantom dark energy prolongs the wavelength that is observed at infinity.

Because we have observed several remarkable features near the Buchdahl limit, it has been fascinating to investigate DES as an ultra-compact object that might produce gravitational echoes. There is an unstable light ring (photon sphere) on $ r=M(2c+1) $ for DES, so we computed the echo time and echo frequency. For DES, there is an imaginary contribution to the echo time. However, the imaginary contributions coming from the integration from the surface to the photon sphere cancel each other. We have found that the phantom field's presence delays the gravitational echoes and makes the frequency becomes bigger in every value of compactness.

Furthermore, the increase of $ c $ corresponds to the increasing echo time or the decreasing echo frequency. Remarkably, the delay of the echoes also corresponds to the effective potential of the perturbed wave equation. We have shown that DES enjoys a more profound (higher) potential well than CDS with the same $ C $. However, we could not portray the potential well of DES in tortoise coordinate since this coordinate is a complex function of $ r $ and we could not find the inverse of $ r_*(r) $ analytically. Hence, in gravitational echoes study, these features obtained from DES compared to that of CDS might be probed in examining dark energy in compact astrophysical objects.

We also have taken $ C=1 $ on DES. This requirement is also the condition to obtain a simple gravastar from CDS. In this limit, it has been found that the interior space-time metric resembles \textcolor{black}{a dS} metric, although with the power of $ c $ on the metric potentials as given in Eq. (\ref{eq:AdSlikemetric}). Obviously, for $ c=1 $, the \textcolor{black}{dS} space-time \textcolor{black}{would be} reproduced. The obtained space-time metric has no singularity as expected for \textcolor{black}{dS} space-time. Unlike the ordinary gravastar from CDS, the dark gravastar pressures are still anisotropic and coordinate-dependent in the interior. Therefore, the radial EoS is still dependent on the coordinate $ r $ and parameters $ M, D$, and $R $, even though the tangential EoS \textcolor{black}{parameter} is coordinate-independent. Since dark gravastar contains an anisotropic structure, we might not assume that there is a thin shell. The exterior space-time is a black hole with a phantom field with an asymptotically flat condition. This dark gravastar possesses a distinct structure as gravastar constructed from CDS. However, dark gravastar similarly possesses no singularity in the space-time metric, pressures, energy density, and charge density.

For future works, it will be interesting to examine the waveform of the gravitational echoes in DES using a numerical method or to use approximation from P{\"o}schl-Teller potential as CDS since they have a similar feature of the effective potential. However, the obstacle to find the tortoise coordinate for the phantom black hole should be resolved first. Moreover, the study of dark gravastar with a thin shell and its thermodynamics will also be possible. In the end, it will be fascinating also to study the gravitational echoes from dark gravastar obtained from DES. Since the future gravitational waves detectors will be available soon, examining ultra-compact DES and dark gravastar will be possibly a new way to probe locally dark energy in compact astrophysical objects.

\begin{acknowledgments}
A. S. is partly supported by DRPM UI's grants No: NKB-1368/UN2.RST/HKP.05.00/2020 and No: NKB-1647/UN2.RST/HKP.05.00/2020. 
\end{acknowledgments}

\appendix

\section{Photon Sphere of Dark Energy Star}
\label{app:photonsphere}
The unstable light ring or photon sphere is an area of space where gravity is extreme in which photons are forced to travel in orbits. For the Schwarzschild black hole, photon sphere is located at $ r=3M $ or  $ r =3r_s/2 $ where $ r_s $ is the Schwarzschild radius. The exterior solution of DES is not similar to the Schwarzschild solution. However, they both have spherical symmetry. The exterior solution of DES is given by Eq. (\ref{eq:extphantommetric}) while the phantom field is given by Eq. (\ref{eq:extphantomfield}).

For a photon traveling at a constant radius $ r $, $ dr=0 $ and $ ds=0 $. It is convenient also to rotate the coordinate system such that $ \theta $ is constant. For instance, we may choose $ \theta =\pi/2 $, so $ d\theta =0 $. Setting these conditions to the exterior metric (\ref{eq:extphantommetric}), we find
\begin{equation}
f^{c} dt^2 = r^2 f^{1-c} \sin^2 \theta d\phi^2, ~~ f= 1-\frac{2m}{r}.
\end{equation}
The above equation the gives
\begin{equation}
\left(\frac{d\phi}{dt} \right)^2 = \frac{f^{2c-1}}{r^2\sin^2\theta}. \label{eq:dphidt}
\end{equation}
Then using the following geodesic equation,
\begin{equation}
\frac{d^2 \gamma^\lambda}{d\tau^2}+\Gamma^{\lambda}_{\mu\nu}\frac{d\gamma^\mu}{d\tau}\frac{d\gamma^\nu}{d\tau} =0,
\end{equation}
we can find 
\begin{equation}
\left(\frac{d\phi}{dt} \right)^2 = \frac{-cf^{2c-1}f'}{r\sin^2\theta (crf' -rf'-2f)}. \label{eq:dphidt1}
\end{equation}
Then using Eqs. (\ref{eq:dphidt}) and (\ref{eq:dphidt1}), one can get
\begin{equation}
r = M(2c+1). \label{eq:photonspheredark}
\end{equation}
This photon sphere radius reduces to the Schwarzschild one when $ D=0 $ or on the other hand, $ c=1 $.
\\
\section{Axial Perturbation}
\label{app:perturbation}
In \cite{UrbanoVeermaeJCAP2019}, it is given that for spherically symmetric metric, 
\begin{equation}
ds^2 = -e^{v(r)}dt^2 + e^{\xi(r)}dr^2 +r^2\left(d\theta^2 +\sin^2\theta d\phi^2 \right),  \label{eq:sphericallysym}
\end{equation}
the single wave equation for a field $ \Psi_{s,l}(r_*,t) $ is given as follows
\begin{equation}
\left[\frac{\partial^2}{\partial t^2}-\frac{\partial^2}{\partial r_*^2} + V_{s,l}(r) \right] \Psi_{s,l}(r_*,t) =0, \label{eq:waveequation}
\end{equation}
where the tortoise coordinate $ dr_* = e^{(\xi-v)/2}dr $ has been used. The general effective radial potential of spin $ s $ is given as
\begin{eqnarray}
V_{s,l}(r) &=& e^{v(r)}\left\{\frac{l(l+1)}{r^2}+\frac{1-s^2}{2re^{\xi(r)}}[v'(r) -\xi'(r)] \right. \nonumber\\
&& - 4 \pi \left[P(r) -\rho(r) \right]    \bigg\} .\label{eq:effectivepot}
\end{eqnarray}
$ P(r),\rho(r) $ are the pressure and energy density of the perfect fluid of the star, respectively. The azimuthal quantum number shall satisfy $ l\geq s $ where $ s $ is the spin of the perturbed field where $ s=0, \pm 1, \pm 2 $ refer to the scalar, vector and tensor field, respectively.

For a gravitational perturbation on the Schwarzschild black hole, one can use the ansatz $ \Psi_{2,l}(r_*,t) = e^{-i\omega_s t}\psi_{2,l}(r) $ to obtain the following wave equation
\begin{equation}
\frac{d^2\psi_{2,l}}{dr^2_*} + [\omega_s^2 - V_{2,l}(r)] \psi_{2,l}=0.
\end{equation}
The effective potential for Schwarzschild black hole is then given by
\begin{equation}
V_{2,l}^{BH} = \frac{1}{r^3}\left(1- \frac{2M}{r} \right)[l(l+1) r - 6M] .
\end{equation}
While for an interior solution (for example CDS), one find
\begin{equation}
V_{2,l}^{Star} =e^{v(r)}\left[\frac{l(l+1)}{r^2} - \frac{6m(r)}{r^3} - 4\pi(P-\rho) \right] .
\end{equation}

\bibliography{apssamp}

\end{document}